\documentstyle[pra,aps,twocolumn,epsf]{revtex}

\begin{document}
\draft
\title{Quantum interference by two temporally distinguishable pulses}
\author{Yoon-Ho Kim, Maria V. Chekhova\thanks{Permanent Address: Department of Physics,
Moscow State University, Moscow, Russia}, Sergei P. Kulik$^*$, Yanhua Shih, Timothy E.
Keller and Morton H. Rubin }
\address{Department of Physics\\ University of Maryland, Baltimore County\\
Baltimore, MD 21250}

\maketitle

\widetext

\begin{abstract}
We report a two-photon interference effect, in which the entangled photon pairs are
generated from two laser pulses well-separated in time. In a single pump pulse case,
interference effects did not occur in our experimental scheme. However, by introducing a
second pump pulse delayed in time $T$, quantum interference was then observed. The
visibility of the interference fringes shows dependence on the delay time $T$ between two
laser pulses. The results are explained in terms of indistinguishability of biphoton
amplitudes which originated from two temporally separated laser pulses.
\end{abstract}


\narrowtext

The superposition principle plays the central role in interference phenomena in quantum
mechanics. In a quantum mechanical picture, interference occurs because there are
indistinguishable ways of an event to occur \cite{Feynman}. Classically, one would not
expect to observe interference from two temporally separated laser pulses. To observe
interference the two pulses, if coherent, must be brought back together in space or
``spread'' by a narrow-band filter. However, it is possible to observe interference
effects for entangled two-photon states generated by two laser pulses which are
well-separated in time.

In this paper, we report a quantum interference experiment in which the interference
occurs between amplitudes of entangled two-photon states generated by two temporally
well-separated laser pulses. The delay between the two pump laser pulses is chosen to be
much greater than the width of the pump pulse and the width of the ``wavepacket"
determined by the spectral filter used in front of the detectors. Therefore, the
``wavepackets" are well distinguishable from single-photon point of view. We first show
why interference effects are not expected for the case of a single pump pulse in our
experimental setup. Then we introduce a second pump pulse delayed in time $T$ and show
how one can recover quantum interference. In this sense, this experiment can be viewed as
a temporal quantum eraser.

The two-photon state in this experiment is generated by spontaneous parametric down
conversion (SPDC). SPDC is a nonlinear optical process in which a higher energy UV pump
photon is converted to a pair of lower energy photons (usually called signal, and idler)
inside a non-centrosymmetric crystal (in this case, $\beta-BaB_2O_4$, called ``BBO"),
when the phase matching condition ($\omega_p=\omega_s+\omega_i$,
$\vec{k}_p=\vec{k}_s+\vec{k}_i$, where the subscripts refer to the pump, signal, and
idler) is satisfied \cite{SPDC}. Signal
\newpage \vspace*{34mm}
\noindent and idler have the same polarization in type I SPDC, and orthogonal
polarization in type II SPDC.

\begin{figure}[tbp]
\centerline{\epsfxsize=2.7in \epsffile{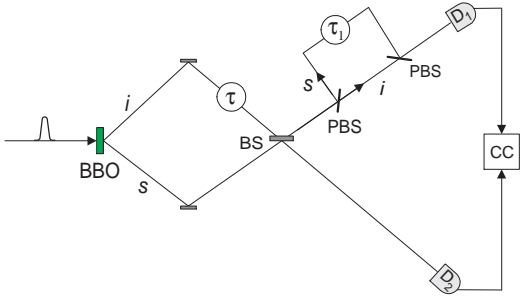}} \caption{Schematic of the experiment.}
\label{fig:scheme}
\end{figure}

\begin{figure}[tbp]
\centerline{\epsfxsize=2.7in \epsffile{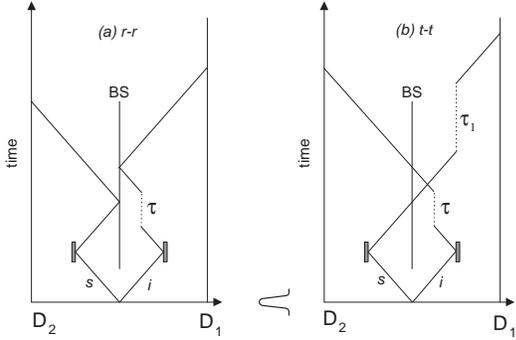}} \caption{Feynman diagrams for the
single pulse case. Two amplitudes are distinguishable since the pump pulse acts as a
clock.} \label{fig:feynman1}
\end{figure}

Let us first consider the case in which a single pump pulse is used for a SPDC process in
the experimental setup shown in Fig. \ref{fig:scheme}. A pair of SPDC photon is fed into
the interferometer. Two photon counting detectors are placed at the two output ports of
the interferometer. The coincidences between the two detectors are recorded. There are
two biphoton amplitudes which could result in coincidence counts for this interferometer:
both signal and idler are (1) transmitted (t-t), (2) reflected (r-r) at the beamsplitter
(BS). If the pump is CW, interference between amplitudes t-t and r-r may occur. Due to
the long coherence length of the pump, the two biphoton amplitudes t-t and r-r may be
indistinguishable, although delays $\tau$ and $\tau_1$ are introduced as shown in Fig.
\ref{fig:scheme} \cite{Pittman96}. However, when a short pulse pump is used, the
interference can never occur. It is, {\em in principle}, possible to know which path (t-t
or r-r) the pair took to contribute a coincidence count. See the Feynman diagrams in Fig.
\ref{fig:feynman1}. One can distinguish t-t and r-r amplitudes by measuring the time
difference between the pulse pump and the detection at one detector since the pump pulse
acts as a clock fixing the origin of the biphoton. In other words, the Feynman
alternatives which originated from a single pulse are distinguishable. It is clear that
no interference will be observed in this case.

In a formal quantum mechanical representation \cite{Keller98}, the field at the detectors
can be written as:
\begin{eqnarray}
E_1^{(+)}(t_1) &=& \frac{1}{\sqrt{2}}\sum_{\omega} e_{\omega} \left( (\hat{e}_x \cdot
\hat{e}_1)a_{1s}(k(\omega))e^{-i \omega(t_1-\tau_1)} \right. \nonumber \\ &+&
\left.i(\hat{e}_y\cdot\hat{e}_1)a_{1i}(k(\omega))e^{-i\omega(t_1-\tau)}\right) \quad ,
\nonumber
\\ E_2^{(+)}(t_2) &=& \frac{1}{\sqrt{2}}\sum_{\omega} e_{\omega} \left(
 (\hat{e}_y\cdot\hat{e}_2)a_{2i}(k(\omega))e^{-i \omega(t_2-\tau)} \right. \nonumber
\\ &+&
\left.i(\hat{e}_x\cdot\hat{e}_2)a_{2s}(k(\omega))e^{(-i\omega t_2)}\right) \quad ,
\end{eqnarray}
where, for example, $a_{1s}(k(\omega))$ is the annihilation operator of a photon arriving
at $D_1$ from the transmitted beam $s$ with wave number $k(\omega)$,
$e_{\omega}=\sqrt{\frac{\hbar \omega}{2 \epsilon_0 V_Q}}$, $V_Q$ is the quantization
volume, and $t_i=T_i-\frac{l_i}{c}$, $i=1,2$ with $l_i$ denoting the optical path length
from the output face of the crystal to detector $i$. $\hat{e}_x$ and $\hat{e}_y$ are the
unit vectors representing the polarization of the photons and $\hat{e}_1$ and $\hat{e}_2$
are the unit vectors representing the orientations of the analyzers placed in front of
the detectors. We can define the scalar projections of the polarizations onto the
detector analyzers as $\hat{e}_x\cdot\hat{e}_i=\cos\theta_i$ and
$\hat{e}_y\cdot\hat{e}_i=\sin\theta_i$, where $i=1,2$. The coincidence counting rate is
given by integrals over the firing times of the two detectors, $T_1, T_2$ respectively
\cite{Glauber}:
\begin{eqnarray}
R_c &\propto& \frac{1}{T} \int_0^T dT_1 dT_2 \langle \Psi |
E_1^{(-)}E_2^{(-)}E_2^{(+)}E_1^{(+)}|\Psi\rangle \nonumber \\ &=& \frac{1}{T}\int_0^T
dT_1 dT_2 \left|\langle0|E_2^{(+)} E_1^{(+)}|\Psi\rangle\right|^2 \quad. \label{RC}
\end{eqnarray}
In our experiment, the pump pulse is linearly polarized and propagating in the
$z$-direction. The central frequency of the pump pulse is $\Omega_p$ and it has an
envelope of arbitrary shape, $\tilde{E}_p$. Then the pump pulse field can be defined by
\begin{eqnarray}
E_p(z,t) &=& e^{-i \Omega_p t} \tilde{E}_p(z,t) \quad, \nonumber \\ \tilde{E}_p(z,t) &=&
\int d\nu_p\bar{E}_p(\nu_p)e^{ik_p(\Omega_p+\nu_p)z-i\nu_pt} \quad.
\end{eqnarray}
The SPDC state vector is given by
\begin{eqnarray}
|\Psi\rangle = |0\rangle + g\sum_{k_s,k_i}\bar{E}_p(\omega_i+\omega_s-\Omega_p) \times
\nonumber \\ a_s^{\dagger} (k_s(\omega_s))a_i^{\dagger}(k_i(\omega_i))|0\rangle \quad,
\end{eqnarray}
where $|0\rangle$ is the vacuum state, and all constants have been gathered in $g$. Here
we consider the degenerate case ($\omega_i=\omega_s$). Let us define
$
t_+ \equiv \frac{1}{2}(t_1+t_2-2\tau), t_{12} \equiv t_1-t_2+\tau.
$
If $t_{12}$ is positive, the idler arrives after the signal. We may think of $t_+$ as the
mean time at which the two-photon wavepacket arrives at the detectors. The {\em biphoton
wavepackets} are ``superposed" in the form:
\begin{eqnarray}
\langle0|E_2^{(+)}E_1^{(+)}|\Psi\rangle = &-&\sin\theta_1\cos\theta_2A(t_+ -
\tau_1,t_{12}-2\tau+\tau_1) \nonumber \\ &+&\cos\theta_1\sin\theta_2A(t_+, -t_{12})
\quad. \label{amplitude}
\end{eqnarray}
In Eq. (\ref{amplitude}) the first term represents the amplitude (wavepacket) where both
signal and idler are transmitted (t-t) at the beamsplitter (BS) and the second term
represents the amplitude (wavepacket) where both signal and idler are reflected (r-r) at
the beamsplitter. Here we have considered only the terms which could result in
coincidence counts. To observe interference, the t-t and the r-r biphoton amplitudes must
be indistinguishable for a given detection time pair ($t_1, t_2$). Indistinguishability
of the two Feynman alternatives implies that it is impossible to tell the biphoton path
from the detection times. It can be shown \cite{Keller98} that the two biphoton
amplitudes cannot be {\em indistinguishable} when we choose a delay $\tau
> \sigma_{pulse}$, the pump pulse width, as in our experiment. Thus the
interference cannot occur in our experimental setup from a single pulse pump.

\begin{figure}[tbp]
\centerline{\epsfxsize=2.7in \epsffile{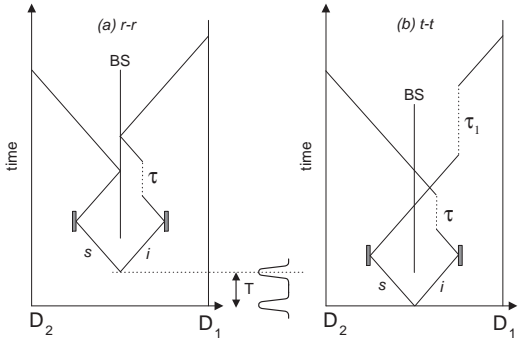}} \caption{Feynman diagrams for the
two-pulse case. r-r (1st pulse) and t-t (2nd pulse) are indistinguishable in terms of the
detector firing times.} \label{fig:feynman2}
\end{figure}

Now, we might ask ourselves, in the case of $\tau > \sigma_{pulse}$, whether it is
possible to ``recover" the quantum interference. This can be accomplished by introducing
a second pulse delayed in time $T$ with $T=\tau>\sigma_{pulse}$. This solution may come
with a serious question: Can interference occur between two temporally distinguishable
pulses? The answer is yes. Fig. \ref{fig:feynman2} shows the Feynman diagrams for this
two-pulse-pump scheme. The complete theoretical treatment can be found in
\cite{Keller98}. The condition for interference is found to be:
\begin{equation}
T=\tau, \quad 2 \tau = \tau_1 \quad.
\end{equation}
When this condition is satisfied, even though (1) the two pump pulses are well
distinguishable in time, (2) detection events for signal or idler are distinguishable,
the r-r amplitude from the first pulse and the t-t amplitude from the second pulse are
indistinguishable with respect to the coincidence detections. This is illustrated in Fig.
\ref{fig:feynman2}. This is a two-photon interference phenomenon between biphoton
amplitudes which originated from two temporally well-separated pulses.

It can be easily shown that the maximum visibility of the interference fringe for two
pulses is $50\%$ by simply using the Feynman diagrams shown in Fig. \ref{fig:feynman2}.
The coincidence counting rate is proportional to $\left|\langle0|E_2^{(+)}
E_1^{(+)}|\Psi\rangle\right|^2$,
\begin{equation}
R_c \propto |A_{1tt}+A_{2tt}+A_{1rr}+A_{2rr}|^2 \quad, \label{prob1}
\end{equation}
where $A_1$ and $A_2$ represent the biphoton amplitudes resulting from the first and the
second pulse pump, respectively. Subscripts $tt$ and $rr$ simply refer to
transmitted-transmitted and reflected-reflected. Considering the non-zero terms of Eq.
(\ref{prob1}) only,
\begin{equation}
R_c \propto 4+A_{2tt}A_{1rr}^* + A_{1rr}A_{2tt}^* \quad , \label{prob2}
\end{equation}
which shows $50\%$ visibility at maximum if $A_{2tt}$ and $A_{1rr}$ overlap. This
deserves further comments. It is usually understood that quantum phenomena show $100\%$
maximum visibility while classical correlation of fields shows $50\%$ maximum visibility
in fourth-order interference experiments \cite{Mandel}. In our case, however, maximum
visibility is $50\%$ although the interference is purely quantum mechanical in nature: it
is due to the quantum entanglement. If we consider multiple pulses ($N>2$) delayed in
time $T$ with respect to one another and include more biphoton amplitudes, maximum
visibility can reach $100\%$ \cite{Keller98}. When $N$ pulse pumps are considered, the
visibility is given by:
\begin{equation}
V(\Delta m, N) = \frac{N-\Delta m}{N} \quad,
\end{equation}
where $\Delta m$ is the difference in pulse numbers of the involved pump pulses. (For
adjacent pulses, $\Delta m = 1$.)

\begin{figure}[tbp]
\centerline{\epsfxsize=2.7in \epsffile{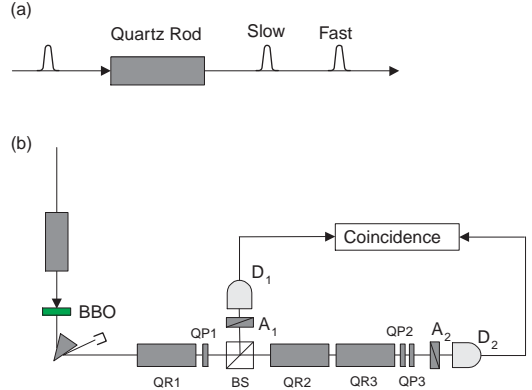}} \caption{(a) Scheme to produce a train
of two pulses. See text for details. (b) Experimental setup. Each detector package
consists of a short focus lens, an interference filter with $10nm$ FWHM, and an avalanche
photo-diode.} \label{fig:type2}
\end{figure}

In our experiment, two temporally separated pump pulses were obtained by transmitting a
single pump pulse through a quartz rod with the optic axis normal to the pump beam and
rotated by 45 deg with respect to the pump polarization. See Fig. \ref{fig:type2}a. Since
the e-ray and o-ray inside the quartz propagate with different group velocities, the
incident single pulse starts to separate into two pulses. The length of the quartz rod
controls the delay between the two output pulses: one polarized in the fast axis
direction and the other polarized in the slow axis direction of the quartz rod. We can
further vary this delay by placing quartz plates after the quartz rod to either make the
delay bigger (optic axis parallel to that of the quartz rod) or smaller (optic axis
perpendicular to that of the quartz rod). The repetition rate of the original pump pulse
is about $90MHz$, so the distance between adjacent pulse is about $11nsec.$ The single
pump pulse has about $140fsec.$ width and central wavelength of $400nm$. The delay ($T$)
between the two pulses is varied from $160\mu m \sim 280\mu m$ (or $533fsec. \sim
933fsec.$), and a coincidence time window of $3ns$ is used. Therefore we can safely say
that the two pulses are temporally well-separated and we only accept biphoton amplitudes
originating from two neighboring pump pulses.

In the experiment, we made use of type II degenerate collinear SPDC
($\lambda_s=\lambda_i= 2 \lambda_p$, where $s$, $i$, and $p$ stands for signal, idler,
and pump respectively). The schematic of the experiment is shown in Fig.
\ref{fig:type2}b. In this scheme, an orthogonal polarized signal-idler photon pair (one
with horizontal polarization, and the other with vertical polarization with respect to
the optic table) propagates in the same direction as the pump. The thickness of the BBO
crystal used in the experiment was $100\mu m$ and the filter bandwidth was chosen to be
$10nm$ ($l_{coh} \simeq \lambda^2/\Delta \lambda \simeq 64 \mu m$). The interferometer
consists of many quartz rods and quartz plates. If the quartz rods (plates) are placed
with optic axis parallel or perpendicular to that of the BBO crystal, they introduce
delays between the signal-idler photon pair. The first delay ($\tau$: QR1 and QP1) is
chosen to be $197\mu m$ (or $657fsec.$) and the second delay ($\tau_1$: QR2, QR3, QP2,
and QP3) is chosen to be $2 \times 197\mu m$. So, here we satisfy one of the two
conditions to observe two-photon interference from two separate pulses, which is $\tau_1
= 2 \tau$. $\tau$ delays idler relative to signal and $\tau_1$ delays signal relative to
idler.

\begin{figure}[tbp]
\centerline{\epsfxsize=2.7in \epsffile{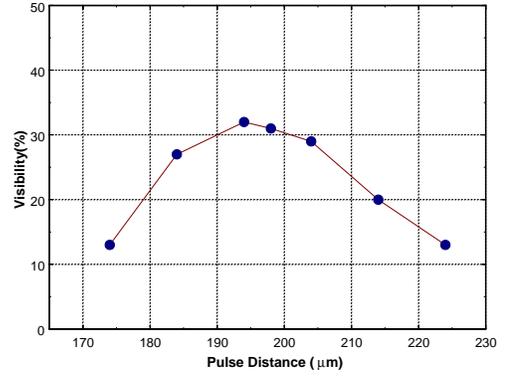}} \caption{Visibility change of the
polarization interference while the delay between two pulses is varied. When $T=\tau$,
maximum visibility was observed.}\label{fig:visibility}
\end{figure}

To demonstrate the two-photon interference effects, we first show the polarization
interference. The analyzer $A_2$ is fixed at $45^o$ and $A_1$ is rotated while recording
the coincidence and single counts. The single counting rate of the detector $D_1$ is
found to be almost constant while the coincidence counts show $\sin^2(\theta_1-\theta_2)$
or $\sin^2(\theta_1+\theta_2)$ modulation depending on the phases introduced when
changing the inter-pulse delay $T$ \cite{Pittman96,polarization,space-time}. We record
the visibility of this polarization interference while varying the delay $T$ between the
two pulses. This is shown in Fig. \ref{fig:visibility}. It is clearly shown that when the
delay $T$ is equal to the interferometer delay $\tau$, the maximum visibility, which in
this case is $33\%$, is observed.

\begin{figure}[tbp]
\centerline{\epsfxsize=2.7in \epsffile{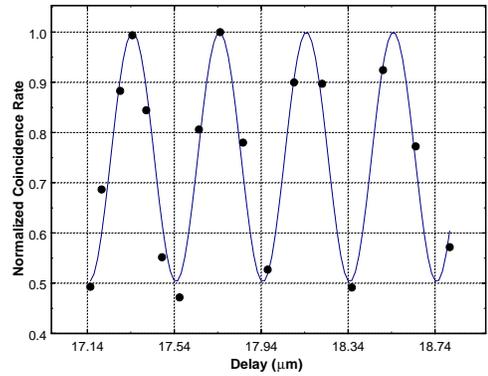}} \caption{Space-time interference
showing dependence on pump phase. The modulation period is $400nm$ as expected from Eq.
(\ref{spacetime}).}\label{fig:pumpmod}
\end{figure}

We also observed the dependence of coincidence counting rate on the phase shift between
the two pump pulses: the space-time interference \cite{space-time}. This is done by
orienting the two polarizers ($A_1$ and $A_2$) at $45^o$ and by placing two additional
quartz plates after QR1. By tilting the quartz plates, we introduce additional phase
delay between the pump pulses. From eqs. (\ref{amplitude}) and (\ref{prob2}), we expect
to observe,
\begin{equation}
R_c \propto 4-2\eta(T)\cos(\Omega_p \phi_p)\quad,\label{spacetime}
\end{equation}
where $\Omega_p$ is the pump frequency, $\phi_p$ is the pump phase, and $\eta(T)$ has a
value of $0 \sim 1$ and reflects the fact that the visibility of the interference fringe
depends on the value of the inter-pulse delay $T$. For the case of maximum visibility,
$\eta(\tau)=1$. The data shown in Fig. \ref{fig:pumpmod} demonstrates dependence on pump
phase change. The modulation period is $400nm$ (pump wavelength) as expected from Eq.
(\ref{spacetime}). Data shown in Figs. \ref{fig:visibility} and \ref{fig:pumpmod} clearly
demonstrate two-photon interference effects between biphoton amplitudes generated from
two temporally separated pump pulses.

In conclusion, we have experimentally demonstrated the two-photon interference between
biphoton amplitudes arising from two temporally well-separated pump pulses. It is
important to note the following. (1) The pump pulse intensity was low enough so that
single counting rates of the detectors were kept much smaller than the pulse repetition
rate: the probability of having one SPDC photon pair per pulse in our experiment is
negligible. Hence the interference cannot be explained as two SPDC photons from two pump
pulses (one SPDC pair from each pulse) interfering at the detectors. The two pump pulses
simply provide two biphoton amplitudes which could result in coincidence counts, and
interference occurs between these two biphoton amplitudes only when they are
indistinguishable. (2) The BBO crystal in our experiment was only $100\mu m$ thick, so
the two-pulse did not exist in the BBO at the same time at any moment. (3) Due to the
delays in the interferometer ($\tau$, $\tau_1$), the signal and idler never met at the
beamsplitter (BS). Therefore, the existence of two-photon interference cannot be viewed
as the interference between the signal and idler photons. These three points again
emphasize the fact that it is the indistinguishability of the Feynman alternatives for
the biphoton amplitudes which is responsible for the quantum interference effects.

This work was supported, in part, by the U.S. Office of Naval Research, and the
National Security Agency. MVC and SPK would like to thank the Russian Foundation for
Basic Research, Grant \# 97-02-17498 for supporting their visit to Univ. of Maryland,
Baltimore County.

\end{document}